
\input harvmac

\noblackbox
\Title{\vbox{\hbox{CTP-TAMU-98/91}}}
{\vbox{\centerline{String Motion in Fivebrane Geometry}}}

\centerline{Ramzi R. Khuri\footnote{$^\dagger$}{%
e-mail address:rrk@phys.tamu.edu, rrk@tamphys.bitnet}\
and HoSeong ~La\footnote{$^*$}{%
e-mail address: hsla@phys.tamu.edu, hsla@tamphys.bitnet}}

\bigskip\centerline{Center for Theoretical Physics}
\centerline{Texas A\&M University}
\centerline{College Station, TX 77843-4242, USA}
\vskip 1in
The classical motion of a test string in the transverse space
of two types of heterotic  fivebrane sources
is fully analyzed, for arbitrary instanton scale size.
The singular case is treated as a special case
and does not arise in the continuous limit of zero instanton size.
We find that the orbits are either circular or open, which is a
solitonic analogy with the motion of an electron around a
magnetic monopole, although the system we consider is quantitatively different.
We emphasize that at long distance this geometry does not satisfy
the inverse square law, but satisfies the inverse cubic law.
If the fivebrane exists in nature and
this structure survives after any  proper compactification,
this last result can be used to test classical ``stringy'' effects.

\Date{12/91} 

\def\la{\lambda}
\def\hf{{1\over 2}} 
\def\half{{\textstyle{1\over 2}}}
\def\d{{\rm d}}
\def\e{{\rm e}}
\def\pa{\partial}

\def\eps{\epsilon}

\font\cmss=cmss10 \font\cmsss=cmss10 scaled 833
\def\IZ{\relax\ifmmode\mathchoice
{\hbox{\cmss Z\kern-.4em Z}}{\hbox{\cmss Z\kern-.4em Z}}
{\lower.9pt\hbox{\cmsss Z\kern-.4em Z}}
{\lower1.2pt\hbox{\cmsss Z\kern-.4em Z}}\else{\cmss Z\kern-.4em Z}\fi}
\def\tr{{\rm tr}}
\def\Tr{{\rm Tr}}
\def\cos{{\rm cos}}
\def\sin{{\rm sin}}
\def\CL{{\cal L}}
\def\CO{{\cal O}}
\def\etp{\e^{2\phi_0}+Q{(r^2+2\la^2)\over(r^2+\la^2)^2}}

\vfill\eject
\newsec{Introduction}

The structures of classical solitonic solutions of string theory have
been actively investigated recently\ref\Fbrev{For recent reviews, see
M.J.~Duff and J.X. Lu, ``A Duality between Strings and Fivebranes,''
Texas A\&M preprint, CTP-TAMU-28/91 (1991);
C.G. Callan, ``Instantons and Solitons in
Heterotic String Theory,'' Princeton preprint, PUPT-1278 (1991);
C.G. Callan, J.A. Harvey and A. Strominger, ``Supersymmetric String Solitons,''
Chicago preprint, EFI-91-66 (1991); and
references therein.}.
Among these solutions
the heterotic fivebrane solution conjectured by Duff\ref\Duf{%
M.J. Duff, Class. Quan. Grav. {\bf 5} (1988) 189\semi
M.J. Duff, in {\it Superworld II}, ed. by A. Zichichi (Plenum, New York, 1990).
}\ and constructed by
Strominger\ref\Stro{A. Strominger, Nucl. Phys. {\bf B343} (1990) 167.}\
is particularly interesting because it is dual
to the fundamental string in the generalized sense of the electric-magnetic
duality\foot{This duality which interchanges
Noether charge (e.g. electric charge)
and topological charge (e.g. monopole charge) is in principle the foundation
for
the Montonen-Olive conjecture\ref\MoOl{C. Montonen and D. Olive, Phys. Lett.
{\bf 72B} (1977) 117.}, which is yet to be confirmed rigorously.}.
Although such a duality does not necessarily imply the existence of the dual
object, e.g. we have not yet found the magnetic monopole, it is worth while to
further investigate the implications of this duality.

In this paper we follow the analogy of the electron-monopole system and shall
study some of the classical motions of a test string around a fivebrane source.
We shall restrict ourselves in this paper to the case in which both string and
fivebrane behave like points in the transverse space of the fivebrane.
For simplicity we also ignore any contribution due to world-sheet fermions.
Nevertheless, even if we include this contribution,
we expect that there will be no qualitative change in the dynamics
because these fermions only couple to the instanton YM background.

We shall consider here both cases of ``gauge'' and ``symmetric'' solutions%
\ref\CaHaSt{C.G. Callan,
J.A. Harvey and A. Strominger, Nucl. Phys. {\bf B359} (1991) 611.}. The
symmetric solution is closely related to the elementary fivebrane solution of
ref.\ref\Dfluone{M.J. Duff and J. X. Lu, Nucl. Phys. {\bf B354} (1991) 141.}.
Though the gauge solution is not an exact
classical background, we still observe various interesting structures, which we
expect will hold in a qualitatively similar way even for the exact fivebrane
background.  Since the symmetric solution  is exact, we can perform an exact
analysis.

In both cases, analogies are made between the string-fivebrane system
and electron orbits around a monopole. For example, in both systems there
exist circular orbits whose discretized radii are governed by the quantization
of the source charge (the instanton charge in the case of the fivebrane and the
magnetic charge in the case of the monopole). Also, both the fivebrane
and the monopole impart an intrinsic angular momentum to the string and the
electron respectively.

Since the dynamics of the
string-fivebrane system is governed by the competition between the attractive
gravitational force and the repulsive force due to the antisymmetric tensor
field, we are led to expect
fundamental differences with General Relativity, in which the latter force
is absent. This expectation is indeed confirmed by our finding that the
long-distance forces obey an inverse cubic law, rather than an inverse
square law, in direct contrast to General Relativity.
This finding could have important
implications for physics according to string theory at scales larger than the
Planck length, provided the fivebrane structure survives compactification.

This paper is organized as follows.
In sect.2 we shall review the derivation of both solutions. In sect.3 we
describe the dynamics generically, for an arbitrary instanton scale size.
In sect.4, we choose a  finite instanton scale size and
to leading order in instanton charge, we study perturbatively the case of the
gauge solution. In sect.5, setting the instanton size to zero, we describe
the symmetric solution case. In both cases we  observe drastic differences
from General Relativity or Newtonian dynamics.
Finally, in sect.6 we shall provide further perspectives.

\newsec{Heterotic Fivebranes}

Let us review the derivations given in refs.\Stro\CaHaSt.
The heterotic fivebrane is a solution to the equations of the supersymmetric
vacuum for the heterotic string
\eqn\suei{\delta\psi_M=\left(\nabla_M-{\textstyle {1\over 4}}H_{MAB}\Gamma^{AB}
\right)\eps=0,}
\eqn\sueii{\delta\lambda=\left(\Gamma^A\pa_A\phi-{\textstyle{1\over 6}}H_{AMC}
\Gamma^{ABC}\right)\eps=0,}
\eqn\sueiii{\delta\chi=F_{AB}\Gamma^{AB}\eps=0,}
where $\psi_M,\ \lambda$ and $\chi$ are the gravitino, dilatino and gaugino,
while
\eqn\anoe{\d H=\alpha' \left(\tr R\wedge R-{\textstyle{1\over 30}}\Tr
        F\wedge F\right).}
In the above we have properly rescaled all the field variables so  that
the string coupling $g_s=\e^{\phi}$ and $\alpha'$ are the only independent
couplings. In the heterotic string $\alpha'$ is proportional to
$\kappa^2/g_{{\rm YM}}$, where $\kappa$ is the gravitational coupling
constant and $g_{{\rm YM}}$ is the YM coupling constant.
Depending on the structure of eq.\anoe, we can have different types of
solutions. First, we shall review the derivation of the so-called ``gauge''
solution and then later the ``symmetric'' solution.

In $(1+9)$-dimension we have Majorana-Weyl fermions, which decompose down to
chiral spinors according to SO(1,9)$\supset$SO(1,5)$\otimes$SO(4) for
the $M^{1,9}\to M^{1,5}\times M^4$ decomposition.
For such spinors the dilatino equation \sueii\ is satisfied by
\eqn\ei{H_{\mu\nu\la}=\pm\eps_{\mu\nu\la\sigma}\pa^\sigma\phi,}
where $\mu,\nu,...$ are indices for the transverse space
$M^4$ and $\phi=\phi(x^\mu)$, while we shall
use indices $a, b,...$ for $M^{1,5}$ below.

The other equations are solved by constant chiral spinors $\eps_{\pm}$ and
\eqn\eii{g_{ab}=\eta_{ab},\ \ \ g_{\mu\nu}=\e^{2\phi}\delta_{\mu\nu}}
such that
\eqn\eiii{\eqalign{\delta\psi_\mu&=\left(\nabla_\mu-{\textstyle{1\over 2}}
\Gamma^{\mu\nu}\pa_\nu\phi\right)\eps_\pm=\pa_\mu\eps_\pm=0,\cr
\delta\psi_a&=\nabla_a\eps_\pm=\pa_a\eps_\pm=0,\cr}}
and
\eqn\eiv{\delta\chi=F^\pm_{\mu\nu}\Gamma^{\mu\nu}\eps_\pm=
-F^\pm_{\mu\nu}\Gamma^{\mu\nu}\eps_\pm=0,}
where eq.\eiv\ is satisfied using an instanton configuration for the
(anti)self-dual YM equation in $M^4$
\eqn\eins{F^\pm_{\mu\nu}=\pm\half\eps_{\mu\nu}^{\ \ \ \rho\sigma}
F_{\rho\sigma}^\pm}
for an
SU(2) subgroup of $E_8\times E_8$ or SO(32). In fact $\phi=\phi(r^2)$ here
(i.e. no angular dependence), where $r^2=\sum(x^\mu)^2$.
With a finite instanton scale size $\lambda$,
\eqn\eexpf{\e^{2\phi}=\e^{2\phi_0}+Q{(r^2+2\lambda^2)\over
(r^2+\lambda^2)^2},}
where $\phi_0$ is the value of the dilaton at spatial infinity and $Q$ is the
charge of the instanton in the unit of $\alpha'$, e.g. $Q=8\alpha'$.

Note that this solution shows a scale symmetry
\eqn\escs{\eqalign{\phi &\to\phi+\ln\sigma,\cr
                r &\to \sigma^{-1}r,\cr
                \la &\to \sigma^{-1}\la,\cr}}
where $\sigma$ is a constant.
The $\la=0$ case is not related to the $\la\neq 0$ case in terms of this scale
symmetry, but it retains a similar scale symmetry without the last property.

Due to such a scale symmetry, when we later describe the dynamics of the
system, particularly in terms of the effective potential,
we need to modify its form slightly in order to
secure the symmetry. Furthermore, there are certain limits of these solutions
which are not related by the above scale symmetry. We thus prefer keeping
$\phi_0$ explicitly instead of fixing its value using the scale symmetry.
For example, $\phi_0=-\infty$ is one extreme limit not related by this scale
symmetry.

In the low energy effective action for the heterotic string the potential for
the dilaton $\phi$ is identically zero if the scale symmetry of the theory is
respected. This scale symmetry is expected to be spontaneously broken to
determine some physically relevant dilaton vacuum expectation value. For any
nontrivial potential for $\phi$ there is always one rather trivial minimum  at
$\phi\to -\infty$, which can be identified as a boundary value of $\phi$, so
that $\phi_0\to -\infty$. Note that at this point
it is clear that the previously mentioned
scale symmetry is spontaneously broken. Though this minimum point is not
really relevant to the low energy physics unless we find another isolated
minimum, it is still worth while to single out this case.

We now have a fivebrane living in $M^{1,5}$ which is a point-like
object in $M^4$. The above solution is called
the ``gauge'' solution\CaHaSt, while
there is another ``symmetric'' fivebrane solution which is exact.

This symmetric
solution can be obtained as follows. Define a generalized connection by
\eqn\genc{\Omega^{AB}_{\pm M}=\omega^{AB}_M\pm H^{AB}_M}
embedded in the SU(2) subgroup and equate it
to the gauge connection $A_\mu$ so that $\d H=0$ and the corresponding
curvature
$R(\Omega_{\pm})$ cancels against the Yang-Mills field strength $F$\ref\Genc{M.
Green and J.H. Schwarz, Phys. Lett. {\bf 151B} (1985) 21\semi R.R. Khuri,
Phys. Lett. {\bf 259B} (1991) 261\semi C. Callan, J. Harvey and A. Strominger,
Nucl. Phys. {\bf B367} (1991) 60.}.
It follows that
\eqn\sdec{R(\Omega_\pm)^{mn}_{\mu\nu}=\mp\half\eps_{\mu\nu}^{\ \ \ \la\sigma}
R(\Omega_{\pm})_{\la\sigma}^{mn}.}
Thus we have a solution
\eqn\symsol{\eqalign{\e^{2\phi} &=\e^{2\phi_0}+{Q\over r^2},\cr
        H_{\mu\nu\la}&=-\eps_{\mu\nu\la\sigma}\pa^\sigma\phi,\cr
        F_{\mu\nu}^{\ \ mn}&=R(\Omega_{-})_{\mu\nu}^{\ \ mn},\cr}}
where both $F$ and $R$ are (anti)self-dual.

This symmetric solution becomes exact since $A_\mu=\Omega_{-\mu}$
implies that all the higher order corrections vanish\ref\BerdR{E.A.
Bergshoeff and M. de Roo, Nucl. Phys. {\bf B328} (1989) 439.}.

\newsec{Generic Case}

For the general case we keep the instanton size $\la$ arbitrary and shall
treat generically both ``gauge'' and ``symmetric'' solutions. For the former
we need only know the solution to leading order in $\alpha'$, but for the
latter we use the exact expression with $\la=0$.
Naively one may associate the symmetric solution with the $\la\to 0$
limit of the gauge solution in such a way that the higher order corrections
vanish as $\la\to 0$. However, we shall later find out that the formal
similarity of these solutions does not necessarily imply the same
limit in the dynamics\foot{ We would like to call the reader's attention
to the fact that in the symmetric solution case the actual instanton size
is not $\la = 0$ but $Q$. Thus the two solutions are not gauge equivalent.}.
This could be anticipated in some sense by observing that the symmetric
solution only makes sense for $\la\gg\sqrt{\alpha'}$\CaHaSt\ so that the limit
$\la\to 0$ does not in fact make sense.

The Lagrangian for a string moving in a given background of massless fields
is given by
\eqn\lag{{\cal L}=\hf\left(\sqrt{-\gamma}\gamma^{ij}\pa_i X^M\pa_j X^N g_{MN}
+ \eps^{ij}\pa_i X^M \pa_j X^N B_{MN}\right)+\cdots,}
where $\gamma_{ij}$ is the worldsheet metric for $(i,j)=(\tau,x)$,
$g_{MN}$ is the ``$\sigma$-model metric'' and $\cdots$ includes the worldsheet
fermion terms, which we ignore for simplicity.
Throughout this paper,
we assume that the string is parallel to one of the fivebrane directions,
i.e. $x$ will be identified as one of the fivebrane coordinates.
Since $B_{MN}$ is only nonzero when both $M$ and $N$ are transverse,
the axion does not contribute to the Lagrangian of the point-like string in the
transverse
space. One can put the test string inside the transverse space, but for our
purposes this is not really necessary, unless we identify the
$(1+3)$-dimensional subspace in order to study the dynamics inside
``space-time''. We will leave this for a future study.

If we substitute the worldsheet constraint equation
$\gamma_{ij}=\pa_i X^M \pa_j X^N g_{MN}$ in \lag,
the relevant Lagrangian for the classical dynamics of the string in the
fivebrane background is simply of the Nambu-Goto type. From \eii\ it follows
that
\eqn\elag{\CL={\sqrt{-\gamma}}=\left[{\dot t}^2-\e^{2\phi}\left\{{\dot r}^2
+r^2\left({\dot\chi}^2+\sin^2\chi({\dot \theta}^2+\sin^2\theta{\dot\varphi}^2)
\right)\right\}\right]^{1/2},}
where $``\cdot"$ is the derivative with respect to the proper time $\tau$.
In contrast to the usual motion of a string in a given background
(e.g. the motion of a cosmic string), we do
not see any derivative along the string direction (i.e. $x$-direction)
because we have identified $x$ with the coordinate outside the transverse space
to make the string point-like here.
Note that the string dynamics is generically different from General Relativity,
in which the geodesic equations are described by, say, the square of the above
Lagrangian.

Though there is no explicit coupling between the string and the antisymmetric
tensor field, the fivebrane geometry originates more or less from the rank
three
tensor $H_{\mu\nu\la}$. We should therefore expect that the dynamics will be
governed by the competition between the attractive force due to gravity and the
repulsive force due to the antisymmetric tensor. In fact we will see later that
while the repulsive force dominates at long distance,
there exists an attractive force region near the center of the
fivebrane. It is noteworthy that this result deviates drastically
from General Relativity.

Using the four-dimensional spherical symmetry of the Lagrangian, we can fix
$\chi$ and $\theta$. For simplicity, we take $\chi=\theta=\pi/2$ so that
the problem reduces effectively to a two-dimensional one described by
polar coordinates $(r,\varphi)$ with a simplified Lagrangian
\eqn\rla{\CL=\left({\dot t}^2-\e^{2\phi}
({\dot r}^2+r^2{\dot\varphi}^2)\right)^{1/2}.}
For time-like geodesics, $\CL=1$\foot{
Note that there is no null geodesic for the Nambu-Goto action induced from the
nonzero string tension Polyakov action because
$\gamma=0$ implies that the induced worldsheet metric $\gamma_{ij}$ is singular
so that it cannot be inverted to define the original worldsheet Polyakov
string action. The story may be different for the zero string tension case.
{}From the pure Nambu-Goto action's point of view, one can certainly allow null
geodesics, but it now takes an infinite amount of ``energy'' to move along
such a null geodesic
so that each point behaves like a ``black hole'' with zero-radius horizon.
}.

This system has two constants of motion along the time-like geodesics.
One is ``energy''
\eqn\eneq{E\equiv{\pa \CL\over\pa {\dot t}}={{\dot t}\over
\CL}={\dot t},}
where $E\geq 1$ can be interpreted as the conserved energy per unit
mass of the string and represents a constant redshift.
The other is ``angular momentum''
\eqn\aneq{L\equiv -{\pa \CL\over\pa {\dot \varphi}}={\e^{2\phi}r^2{\dot
\varphi}
\over\CL}=\e^{2\phi}r^2{\dot\varphi}.}
$L$ can be interpreted as the conserved angular momentum of the string per
unit mass and may be rewritten in the form $L=L_0+Ql$.
Recall that the dependence of $\e^{2\phi}$ on the instanton solution eq.\eexpf\
tells us that the angular momentum is related to the instanton charge. This is
analogous to the situation of the electron-monopole system, in which the
angular
momentum of the electron shifts due to the monopole charge. Thus
the first term $L_0$ represents the free angular momentum, while the second
term
$Ql$, which is proportional to $Q$, represents  the intrinsic angular momentum
provided by the fivebrane. In this case, however, the motion of the
string is restricted to its initial plane.
Indeed we shall find that some of the characteristic motions of the test
string around the fivebrane are analogous to the motions of an electron around
a monopole source\ref\Monosca{D.G. Boulware, L.S. Brown, R.N. Cahn, S.D.
Ellis and C. Lee, Phys. Rev. D {\bf 14} (1976) 2708.}.

One more comment on the energy: $E$ does not in fact scale as an
energy with respect to the length scale for $r$
but $\tilde{E}\equiv E\e^{-\phi_0}$ does. In order to define the ``effective
potential'' $V$ we rewrite the geodesic condition $\CL=1$ in the form
\eqn\evef{{\dot r}^2+V^2={\tilde E}^2.}
For $\phi_0\to -\infty$, the scale symmetry is missing as we explained before
and we can define the effective potential from $\dot{r}^2+V^2 = E^2$ now, where
$V^2$ can be easily determined. However, the behavior of $V^2$ seems to be
rather unrealistic in most cases except the symmetric solution case, where for
$a^2>0$ $V^2$ has a minimum at $r^2=0$ and is then monotonically increasing.
This can be interpreted as a ``confinement'' of the test string at the
fivebrane source point.

{}From eqs.\eneq\aneq\ and the geodesic condition we get
\eqn\dphdt{{\dot\varphi}={L\over r^2\left[\etp\right]}}
and
\eqn\eret{\dot{r}^2={E^2-1\over \etp} -{L^2\over r^2\left[\etp\right]^2}.}

{}From eq.\eret\ we can read off the effective potential
\eqn\evv{V^2=E^2\e^{-2\phi_0}-{E^2-1\over\etp}
+{L^2\over r^2\left[\etp\right]^2}.}
The turning points, if any, are found by setting $V=\tilde{E}$
(i.e. $\dot{r}=0$).

If $E^2=1$, we have $\dot{r}=0=\dot{\varphi}$ and $L=0$ and the string remains
stationary with no force acting on it. This is analogous to
the electron-monopole case in the sense that if the electron is stationary,
there is no force acting on it due to the monopole.

To determine the turning points for $E^2>1$ we must solve the following
cubic equation for $x\equiv r^2$:
\eqn\cubeq{x^3+(2\la^2-\tilde{a}^2)x^2+\la^2(\la^2-2\tilde{a}^2)x
-\la^4(\tilde{a}^2+\tilde{Q})=0,}
where $\tilde{a}=a\e^{-\phi_0}$, $a^2\equiv{L^2\over E^2-1}-Q$ and
$\tilde{Q}=Q\e^{-2\phi_0}$.
One can easily see that this equation has at least one root at
$\sqrt{x}=r_{{\rm min}}\geq 0$.
In fact there is exactly one root in the physical region $x\geq 0$.
The physical region is therefore restricted to $\sqrt{x}\geq r_{{\rm min}}$.
In contrast to General Relativity, there are no bound orbits,
except for a trivial unstable circular orbit at $r=r_{{\rm min}}$.
This can be seen as the result of
the dominance of the long range repulsive force over the
long range attractive force of standard General Relativity.

The motion from the observer's point of view can be described by
$\d\varphi/\d t=\dot{\varphi}/\dot{t}=\dot{\varphi}/E$
and $\d r/\d t=\dot{r}/E$. The asymptotic behavior of $\dot{\varphi}$ and
$\dot{r}$ is as follows: $\dot{\varphi}\to 0,\ \dot{r}\to{\rm const.}$ as $r\to
\infty$ and $\dot{\varphi}\to {\rm const.},\ \dot{r}\to 0$ as $r\to
r_{{\rm min}}$.

The orbits can be classified according to the
signs of $\dot{r}$, $\dot{\varphi}$ or $\d r/\d t$, $\d\varphi/\d t$;
but it is always useful to take an analogy with the Keplerian analysis of
Newtonian dynamics to describe the orbits. In this way we can also obtain a
reference-frame independent classification of the orbits, although the
situation here is quite different from that of the Keplerian orbits.
As usual we introduce the variable $u\equiv 1/r$ for convenience.

For this purpose we first write down the orbit equation
\eqn\eqor{\left({\d r\over\d\varphi}\right)^2={E^2-1\over L^2}r^4
\left[\etp\right]-r^2.}
Note that, setting  $\d r/\d\varphi=0$, we recover the cubic equation
eq.\cubeq\ for $L\neq 0$. In terms of $u$ we obtain
\eqn\equru{\left({\d u\over\d\varphi}\right)^2={E^2-1\over L^2}
\left[\e^{2\phi_0}+Q {u^2+2\la^2 u^4\over (1+\la^2 u^2)^2}\right]-u^2.}
Later we shall find out that the solutions of this differential equation are
in general not conic.

To see the competition between the attractive and repulsive forces we must
compute the acceleration of the test string. The radial component is given by
\eqn\acr{a_r=Q{r^2+3\la^2\over r (r^2+\la^2)^3\left[\etp\right]^3}
\left[(E^2-1)r^2\left\{\etp\right\}-2L^2\right]}
and the angular component is
\eqn\acph{a_{\varphi}=2QL\dot{r} {r^2+3\la^2\over \left[\etp\right]^2
(r^2+\la^2)^3},}
where $\dot{r}$ is given by eq.\eret. If we divide
by $E^2$, we can obtain the components of the acceleration measured by
a far away observer. Needless to say, if $Q$ vanishes, the acceleration
vanishes. It is especially noteworthy that $a_r\propto Q r^{-3}$ as
$r\to\infty$, as opposed to the usual long range inverse square law!
We instead have an inverse cubic law\foot{Such an inverse cubic law appears in
the electron-monopole case, but only in terms of the modified position vector.
For more detail, see ref.\Monosca.}.
If this structure survives after compactification to (1+3)-dimensional
space-time, it could be used to test classical ``stringy''
effects on the dynamics.

Based on the sign of the radial component $a_r$, we can tell whether the force
is attractive or repulsive. But, the characteristic features of the force
depend on the angular momentum $L$ since the attractive force can be
generated only if there is angular motion. Generically, there are two regions.
In one region (short distance)
the attractive force dominates and in another region (long distance)
the repulsive force dominates. The attractive force region shrinks as the
scale size of the instanton becomes smaller.
If $L=0$, the attractive force region disappears. For $L\neq 0$, there are
always two regions.

\newsec{Gauge Solution}

Since the gauge solution is not exact, the previous analysis is in fact
overdone
and is only meaningful to leading order in $\alpha'$.
If we solve the cubic equation \cubeq\ perturbatively, we find the turning
point
at
\eqn\turpoi{
r_{{\rm min}}
={L\over\sqrt{E^2-1}\e^{\phi_0}}\left[
1-\half Q\e^{2\phi_0}{{L^2\over (E^2-1)\e^{2\phi_0}}+2\la^2\over
\left\{ {L^2\over (E^2-1)\e^{2\phi_0}}+\la^2\right\}^2}\right]+\CO(Q^2)
.}
Note that $r_{{\rm min}}> \left| a\e^{-\phi_0}\right|$ for $\la\neq 0$.
Thus there is always a turning point in this case.

Now to find out the structure of the orbits let us reexamine the effective
potential eq.\evv\ only up to $\CO(Q)$:
\eqn\gaevv{V^2=\e^{-2\phi_0}+{L^2\over r^2}\e^{-4\phi_0}
+Q\e^{-4\phi_0}{r^2+2\la^2\over (r^2+\la^2)^2}\left[(E^2-1)-{2L^2\over r^2}
\e^{-2\phi_0}\right] +\CO(Q^2).}
Note that this potential does not have any critical points but is monotonically
decreasing. As $r\to\infty$, $V^2\to\e^{-2\phi_0}$ regardlessly of the angular
momentum.
In the $r\to 0$ limit the situation depends on the angular momentum.
If $L\neq 0$, $V^2\to\infty$ as $r\to 0$, but if $L=0$,
$V^2\to\e^{-2\phi_0}+2Q\e^{-4\phi_0}(E^2-1)/\la^2$ as $r\to 0$.
Since $\tilde{E}^2\geq\e^{-2\phi_0}$, all the orbits are open
except for a trivial circular orbit.

Since the solution we have at hand is valid only for
$\lambda\gg\sqrt{\alpha'}$, it is not surprising that in the limit $\la\to 0$
we do not recover the limit of the symmetric solution case,
which will be studied in the next section. In other words, the double limit
$r\to 0$ and $\la\to 0$ is not order independent.

We can attempt to solve the equation of motion eq.\equru\ perturbatively.
To order $Q$ we obtain
\eqn\eusol{u={\sqrt{E^2-1}\e^{\phi_0}\over L}\cos(\varphi-\varphi_0)+Qv,}
where $v$ satisfies the differential equation
\eqn\evdef{\eqalign{\sin(\varphi-\varphi_0){\d v\over \d\varphi}&=
v\cos(\varphi-\varphi_0)\cr &-\half{\sqrt{E^2-1}\over L\e^{\phi_0}}
{ {E^2-1\over L^2}\e^{2\phi_0}\cos^2(\varphi-\varphi_0) + 2\la^2\left\{
{E^2-1\over L^2}\e^{2\phi_0}\cos^2(\varphi-\varphi_0)\right\}^2 \over
\left\{1+\la^2{E^2-1\over
L^2}\e^{2\phi_0}\cos^2(\varphi-\varphi_0)\right\}^2}.\cr}}
Note that the leading order term in eq. \eusol\ describes simply a straight
line so that it is clear that with the correction the solution does not show
conic motion.

{}From eqs.\acr\acph\ we can compute the components of the acceleration up to
$\CO(Q^2)$ as follows:
\eqn\gauar{a_r=Q{r(r^2+3\la^2) A_2\over \e^{4\phi_0}(r^2+\la^2)^3}
        -2Q^2{r(r^2+2\la^2)(r^2+3\la^2) A_3\over
        \e^{6\phi_0}(r^2+\la^2)^5}+\CO(Q^3),    }
and
\eqn\gauap{a_{\varphi}
=\pm Q{2L(r^2+3\la^2)\over      \e^{5\phi_0}(r^2+\la^2)^3}A_1^{1/2}
        \mp Q^2{L(r^2+2\la^2)(r^2+3\la^2)\over
        \e^{6\phi_0}(r^2+\la^2)^5}\left[4+{A_2\over \e^{\phi_0}A_1^{1/2}}
\right] +\CO(Q^3), }
where $\pm$ is determined by the sign of $\dot{r}$ and
$A_n\equiv (E^2-1)-n{L^2\over r^2\e^{2\phi_0}}$.

The radial position $r_a$ at which $a_r=0$ is given by
\eqn\eraa{r_a={\sqrt{2}L\over\sqrt{E^2-1}\e^{\phi_0}}\left[1+\half Q (E^2-1)
{ {2L^2\over (E^2-1)\e^{2\phi_0}}+2\la^2\over\left\{
{2L^2\over (E^2-1)\e^{2\phi_0}}+\la^2\right\}^2}\right]+\CO(Q^2).}
Note that $r_a>r_{{\rm min}}$ and $a_r>0\ (a_r<0)$ for $r>r_a\ (r<r_a)$.

Using the ratio
\eqn\accrat{{a_\varphi\over a_r}=\pm{2L\over r\e^{\phi_0}}{A_1^{1/2}\over A_2}
\left[1+\tilde{Q}{(r^2+2\la^2)\over (r^2+\la^2)^2}\left(2{A_3\over A_2}
-{1\over 2L}{A_2\over A_1}\right)\right]+\CO(Q^2),   }
we have
\eqn\forcrat{\left|{a_\varphi\over a_r}\right| > \left|{rd\varphi\over dr}
\right|,}
from which it follows that the trajectory always bends concave
(i.e. inward) with respect to the origin.

There are no closed orbits except a trivial unstable circular orbit with
radius $r=r_{{\rm min}}$. If the radial component of the velocity of the
string is initially directed towards the
fivebrane, it will spiral in to the turning point and then spiral
away to infinity. If the string is directed away from the fivebrane, it will
spiral away to infinity in a concave trajectory.

\newsec{Symmetric Solution}

For the symmetric solution we now keep the exact form of the generic case
and set $\la=0$.  For $L=0$ we have radial motion, and there is no turning
point, as in the elementary fivebrane case%
\ref\DKL{M.J. Duff, R.R. Khuri and J.X. Lu, Texas A\&M preprint, CTP-TAMU
89/91 (1991).}. The force is always repulsive and the
attractive force region shrinks down to zero.

For $L\neq 0$ the cubic equation
reduces to a linear equation leading  to a turning point at
\eqn\sytur{r_{{\rm min}}=\tilde{a},}
provided that $L^2> (E^2-1)Q$. If $L^2\leq (E^2-1)Q$, then $r_{{\rm min}}=0$.

If initially the string is at $r_{{\rm min}}$ with $\dot{r}=0$, for constant
$\dot{\varphi}$ there is an unstable circular orbit. For given $E$ and
$L$ the size of this circular orbit depends only on the instanton charge $Q$
of the fivebrane source. Since $Q$ is quantized in terms of $\alpha'$,
$r_{{\rm min}}$ is discretized accordingly. This is also analogous to the
circular orbit of an electron around a monopole, in which the radius is
discretized by the monopole charge.

In the symmetric solution case the effective potential can be simplified to
\eqn\syevv{V^2=E^2\e^{-2\phi_0}-{E^2-1\over \e^{2\phi_0}+{Q\over r^2}}
+{L^2\over r^2\left[\e^{2\phi_0}+{Q\over r^2}\right]^2}.}
In the domain $r\geq r_{{\rm min}}$ the potential does not have any
critical points but is monotonically
decreasing. As $r\to\infty$, $V^2\to\e^{-2\phi_0}$ regardlessly of the angular
momentum, as in the gauge solution. This implies that the long distance
behavior is similar to the gauge solution case. In other words, we cannot
distinguish the two cases by looking at the long-distance behaviour.
The $r\to 0$ limit of $V^2$, however, does not depend on the angular momentum
in this case, so that $V^2\to E^2\e^{-2\phi_0}$ as $r\to 0$.

We now rewrite the orbit equation as
\eqn\esqor{\left({\d r\over\d\varphi}\right)^2={E^2-1\over L^2}r^4
\left[\e^{2\phi_0}+{Q\over r^2}\right]-r^2,}
which can be simplified further using $a$ as
\eqn\essqo{\left({\d r\over\d\varphi}\right)^2={r^2\over a^2+Q}
\left(r^2\e^{2\phi_0}-a^2\right).}
Again in analogy with the Keplerian analysis of Newtonian dynamics we
reparametrize by $u\equiv 1/r$ and obtain
\eqn\eskepl{\left({\d u\over\d\varphi}\right)^2={1\over a^2+Q}
\left(\e^{2\phi_0}-a^2 u^2\right).}
This equation of motion can be easily solved with the result
\eqna\symeom
$$\eqalignno{u&={1\over r}
={1\over\tilde{a}}\cos\left(\omega(\varphi-\varphi_0)\right)
\qquad\hbox{for } a^2> 0,&\symeom a\cr
u&={1\over r}
={1\over i\tilde{a}}\cosh\left(\omega(\varphi-\varphi_0)\right)
\qquad\hbox{for } a^2\leq 0,&\symeom b\cr}$$
where $\omega^2=\left|a^2\right|/(a^2+Q)$ and $\tilde{a}=a\e^{-\phi_0}$ as
before. The solution for the $a^2=0$ case
$u=\pm e^{\phi_0}Q^{-1/2}\left(\varphi-\varphi_\infty\right)$ arises as
a limiting case of both $\symeom{a}$ and $\symeom{b}$.
Note that in both cases the orbits are not conic! Thus we do not recover
any of the orbits of Newtonian dynamics. Even if we compare the above orbits
with those of Schwarzschild geometry, it is easy to see that there are no
similarities.

The above solutions $\symeom{a}$ and $\symeom{b}$ correspond to two
qualitatively different classes of orbits determined from the initial
conditions. For both cases, it follows from the monotonicity of the effective
potential in the range $r\geq r_{{\rm min}}$ that if the test string is
directed initially away from the fivebrane (i.e. the radial component of the
initial velocity is positive), then it will spiral away to infinity.
If the string is directed towards the fivebrane, then in the two cases the
motion is as follows: If $L^2\leq (E^2-1)Q$ ($a^2\leq 0$) as in
eq. $\symeom{b}$, the string spirals into the fivebrane in an infinite amount
of proper time, thus never observing a singularity (the special case
of $L=0$ for the elementary fivebrane was discussed in \DKL).
If $L^2> (E^2-1)Q $ ($a^2> 0$) as in eq. $\symeom{a}$, there is a point of
closest approach $r_{{\rm min}}=\tilde a$, after which the string swings back
to infinity.
A geometrical way to distinguish the two cases from the initial conditions
is to draw a $3$-dimensional cone in the transverse four space with vertex
at the string, axis along the radial direction
and half-angle ${\rm Arctan}(\sqrt{Q}/r)$. If the velocity vector (either
proper or coordinate) lies within the cone then $a^2\leq 0$. Otherwise,
$a^2> 0$.

As in the previous two sections, we compute the components of the acceleration
of the test string in order to study the competition between
the attractive and repulsive forces. The radial component is given by
\eqn\ascr{a_r={Q\over r^5 \left[\e^{2\phi_0}+{Q\over r^2}\right]^3}
\left[(E^2-1)r^2\left\{\e^{2\phi_0}+{Q\over r^2}\right\}-2L^2\right]}
and the angular component is
\eqn\ascph{a_{\varphi}=\pm
{2QL \over r^5\left[\e^{2\phi_0}+{Q\over r^2}\right]^3}
\left[(E^2-1)r^2\left(\e^{2\phi_0}+{Q\over r^2}\right)-L^2\right]^{1/2},}
where the sign is chosen according to the sign of $\dot{r}$.
Again dividing by $E^2$, we can obtain the components of the
acceleration measured by a distant observer.

Again we have $a_r>0$ ($a_r<0$) if $r>r_a$ ($r<r_a$),
where here
$r_a^2=\left({L^2\over E^2-1}+a^2\right)\e^{-2\phi_0} > r^2_{{\rm min}}$
for $L>0$.
Recall that for radial motion $r_a=0$ and $a_r>0$ always and the ``force'' is
always outward, i.e. repulsive. Usually the angular momentum signals the
existence of an outward force region,
but here we have an extra inward force region
$r_a>r>r_{{\rm min}}$ because of the nonzero angular momentum, as discussed
before. The source of this attractive force is the instanton at
the center. This phenomenon is not completely unknown.
For example, in the electron-monopole case the angular motion of the
electron generates a current which interacts with the monopole charge to
generate an extra force.

It is easy to see from the acceleration that for both types of orbits we have
\eqn\forceratio{\left|{a_\varphi\over a_r}\right| > \left|{rd\varphi\over dr}
\right|,}
from which it follows that the trajectory always bends concave (i.e. inward)
with respect to the origin.
Another way to see this is to note that for the $L^2> (E^2-1)Q$ ($a^2> 0$)
case, $\Delta\varphi=\pi(\e^{2\phi_0}+Q/a^2)^{1/2} > \pi$,
where $\Delta\varphi$ is the
total angular deviation. In fact, one can determine from the initial conditions
the number of loops the string makes around the fivebrane before heading off
to infinity. $n$ loops means
$2\pi n < \Delta\varphi \leq 2\pi ( n+1)$
which is equivalent to
\eqn\loops{\left( 1- {1\over 4n^2}\right)\left( {L^2\over E^2-1}\right)
 < Q \leq
\left(1- {1\over 4(n+1)^2}\right)\left( {L^2\over E^2-1}\right).}
Note that as $n\to\infty$, $a\to 0$ and the string spirals towards the
fivebrane. The ``swingshot''
orbits for $a^2> 0$ are analogous to electron-monopole
orbits in which the electron swings around the monopole before heading off to
infinity.

\newsec{Discussion}

For nonzero angular momentum $L$ the motion of the test string around a generic
heterotic fivebrane source in the transverse space is either circular or
open. The open orbits spiral in to the turning point, then spiral away to
infinity. The radius of the circular orbit is governed by the quantized
instanton charge $Q$ and is discretized accordingly, in analogy with the
radius of the circular orbit of an electron around a monopole, which is
governed by the quantized monopole charge. Other analogies with the
electron-monopole system include the shifting of the angular momentum of
the string due to the fivebrane, and the existence of open swingshot orbits.

Furthermore, the open orbits are not conic. This signifies that the string
motion around the fivebrane differs fundamentally from previously studied
orbits
in gravitational theory. In fact, the radial component of the acceleration goes
as $Q r^{-3}$ asymptotically as $r\to\infty$. This implies that the force
does not satisfy the inverse square law, in direct contrast with
General Relativity. The source of this departure lies partially
in the presence of the antisymmetric tensor
in the string-fivebrane system, which generates an
additional repulsive force which dominates the dynamics at long distances and
precludes the existence of stable bounded orbits.

Although our analysis has been done in the transverse space of the fivebrane
only, the implications of our findings could be significant.
If these structures survive
compactification, one might be able to formulate an interesting test for
string theory. Either there is no remnant of the fivebrane after
compactification if we fail to observe these new structures, or string theory
should lead to structures different from those of Newtonian dynamics at
distances longer than the Planck scale, yet at a still sufficiently short
scale (otherwise, this structure may not be observable anyhow). It is
therefore important to seriously investigate the compactifications of
the fivebrane solutions of ten-dimensional string theory.

\bigbreak\bigskip\bigskip\centerline{{\bf Acknowledgements}}\nobreak

\par\vskip.3truein

The authors would like to thank M. Duff for discussions.

This work was supported in part by NSF grant PHY89-07887 and World Laboratory
Fellowships.


%
\listrefs
\vfill\eject
\bye